\documentclass[twocolumn]{aastex7}

\usepackage{amsmath}
\usepackage{multirow}
\usepackage{booktabs}
\usepackage{caption} 
\usepackage{float}
\shorttitle{Ayubinia et al.}
\shortauthors{Scaling Relations of the Torus}

\begin{document}

\title{{Scaling Relations of the Dusty Torus with luminosity and the broad-line region}}

\author[orcid=0009-0002-9526-5820]{Ashraf Ayubinia}
\affiliation{Department of Physics \& Astronomy, Seoul National University, Seoul 08826, Republic of Korea}
\email[]{}

\author[orcid=0000-0002-8055-5465]{Jong-hak Woo}
\affiliation{Department of Physics \& Astronomy, Seoul National University, Seoul 08826, Republic of Korea}
\email[show]{jhwoo@snu.ac.kr} 

\author[orcid=0000-0002-2052-6400]{Shu Wang}
\affiliation{Department of Physics \& Astronomy, Seoul National University, Seoul 08826, Republic of Korea}
\email[]{} 

\author[orcid=0000-0001-9957-6349]{Amit Kumar Mandal}
\affiliation{Department of Physics \& Astronomy, Seoul National University, Seoul 08826, Republic of Korea}
\affiliation{Center for Theoretical Physics of the Polish Academy of Sciences, Al. Lotnik\'{o}w 32/46, 02-668 Warsaw, Poland}
\email[]{} 

\author[orcid=0000-0002-4704-3230]{Donghoon Son}
\affiliation{Department of Physics \& Astronomy, Seoul National University, Seoul 08826, Republic of Korea}
\email[]{} 

\begin{abstract}
We measure and compare the size of the dusty torus with active galactic nucleus (AGN) luminosity and the size of the broad-line region (BLR), using a sample of \textcolor{black}{182} AGNs with the best H$\beta$ lag measurements. After correcting for accretion-disk contamination, torus sizes are determined from the time lags of the Wide-field Infrared Survey Explorer W1 and W2 band light curves relative to the optical band variability based on the interpolated cross-correlation function (ICCF) analysis and the Multiple and Inhomogeneous Component Analysis. We find that the torus size from the W1-band (W2-band) tightly correlates with the 5100\,\AA\ continuum luminosity with an intrinsic scatter of 0.15-0.16 dex and the best-fit slope of $0.35 \pm 0.03$ ($0.33 \pm 0.03$), which is clearly shallower than the expected 0.5 slope from the sublimation radius-luminosity relation. We find a moderate negative trend that higher Eddington AGNs tend to have smaller torus sizes than expected from the best-fit, suggesting the Eddington ratio plays a role in flattening the torus size-luminosity relation. By comparing the torus size with the H$\beta$ reverberation time lag for a subsample of 67 AGNs, we find that the torus size is a factor of $\sim 10$ and $\sim 14$ larger than the BLR size, respectively for W1 and W2 bands. The torus size based on the W1 (W2) band correlates with the BLR size with the best-fit slope of $1.28 \pm 0.16$ ($1.10 \pm 0.15$), which is comparable but slightly steeper than a linear correlation.
\end{abstract}\

\section{Introduction}
\setcounter{footnote}{0}
The unified model of active galactic nuclei (AGNs) \citep[e.g.,][]{Antonucci1993,Urry1995,Netzer2015} postulates a central supermassive black hole (SMBH) surrounded by an accretion disk, broad-line region (BLR), and an obscuring dusty torus. The accretion disk emits UV and optical radiation, ionizing the surrounding gas in the BLR. Enveloping this region, the dusty torus reprocesses radiation into infrared (IR) emission, with its inner boundary located at the dust sublimation radius \citep[e.g.,][]{Edelson1986,Elvis1994,Honig2017}.

A distinctive bump at $\sim 3-5\ \mu m$ in the AGN spectral energy distribution (SED) arises from the thermal emission of hot dust near the sublimation radius where dust grains are heated to $\sim$1500 K and sublimate due to intense UV/optical radiation from the accretion disk \cite[]{Barvainis1987}. Under the assumption of thermal equilibrium, the sublimation radius provides a robust estimate of the inner boundary of the torus, as demonstrated by \cite{Barvainis1987}, scaling with the accretion disk luminosity as $R_{\rm sub} \propto L^{0.5}$. This indicates that stronger radiation shifts the sublimation radius outward in more luminous AGNs. Beyond this region, cooler dust extends outward, reprocessing the absorbed radiation into longer IR wavelengths and significantly contributing to the mid-infrared (MIR) emission observed in AGNs.

It is extremely challenging to spatially resolve the dusty torus due to its compact size \citep[$\sim$ subparsec to a few parsecs;][]{Minezaki2004, Packham2007, Mandal2024} and the contamination from the accretion disk in type 1 AGNs, which hinders the isolation of torus emission despite having a direct view of the central region. MIR data have revealed warm dust on 3-30 mas scales ($z \sim 0.01$; 30 mas $\sim$6 pc) \citep[][]{Jaffe2004,Tristram2009,Kishimoto2011,Burtscher2013}, while near-IR (NIR) interferometry achieves sub-mas resolution, inferring ring radii from sub-pc to several pc \citep[][]{Weigelt2012,Leftley2021,Kishimoto2022,Gravity2020,Gravity2024}. These findings confirm that interferometric dust sizes closely follow size-luminosity trends from dust sublimation models \citep[e.g.,][]{Kishimoto2007,Kishimoto2009,Gravity2024}, while such studies remain limited to a small set of nearby, bright AGNs.

Alternatively, torus size can be determined via reverberation mapping (RM), which measures light-travel time from the accretion disk to surrounding dust \cite[]{Clavel1989}. Early NIR RM studies validated the method by reporting K-band dust lags in nearby AGNs, establishing a tight torus size-luminosity relation consistent with dust sublimation physics \citep[e.g.,][]{Suganuma2006,Koshida2009,Koshida2014,Mandal2021}. Studies of more luminous AGNs confirmed the trend but found a shallower slope than the expected 0.5 \cite[e.g.,][]{Minezaki2019}, suggesting the underlying physics is not fully understood. Compared to interferometric measurements in the same band, RM-based dust radii are typically smaller by a factor of $\sim$2, likely due to differences between response-weighted and flux-weighted radii. While interferometry captures cooler dust at larger radii, RM traces variability in hotter inner dust, making it a better proxy for the torus inner edge \citep[][]{Kishimoto2011,Koshida2014}. 

While the number of AGNs with NIR RM measurements remains small, recent studies extended RM to the MIR using all-sky data from the Wide-field Infrared Survey Explorer (WISE), enabling larger AGN samples. These WISE-based studies reveal a torus size-luminosity relation consistent with NIR RM,  reporting a shallower slope than 0.5 \citep[e.g.,][]{Lyu2019,Yang2020,Chen2023,Mandal2024}.

Using the RM technique, the BLR size can be constrained via time delays between ionizing continuum and broad-line variations \citep[e.g.,][]{Du2015,Grier2017,Cho2023,Woo2024}. Comparing BLR and torus sizes offers key insights into AGN structure.  \citet{Lyu2019} reported that the dust torus lies just beyond the BLR with the torus being larger by a factor of $\sim$4.3 based on 15 PG quasars. Building on this, \citet{Chen2023} analyzed 78 AGNs and showed that the torus is larger than the H$\beta$ BLR by a factor of $\sim$9.2 in W1 ($\sim$11.2 in W2), reporting a linear torus size-H$\beta$ BLR relation. \citet[][]{Mandal2024} also reported that torus size is typically larger than the BLR size by a factor of $\sim$9.5  in W1 ($\sim$15.1 in W2). 

Expanding upon recent progress and the increasing number of AGNs with BLR reverberation mapping, we investigate the torus size scaling relations with AGN luminosity and the BLR size, incorporating several improvements. First, we extend the temporal baseline by incorporating both optical and MIR light curves, using the complete WISE dataset up to its conclusion in August 2024. Second, we use the uniformly measured $\rm H\beta$ lag from \cite{Wang2024} to investigate the torus and BLR sizes. Third, we include 32 luminous AGNs from the Seoul National University AGN Monitoring Project \cite[SAMP,][]{Woo2019,Woo2024}, increasing both the sample size and the representation of high-luminosity AGNs in the size-luminosity relation. 
Fourth, we correct MIR light curves for accretion disk contamination, which can otherwise lead to underestimated torus size as highlighted by \citet[][]{Mandal2024}, 
to enhance the reliability of dust lag measurements. In this paper we describe the sample selection in Section \ref{sec:sample} and the data analysis in Section \ref{sec:data_analysis}. Results and discussion are presented in Sections \ref{sec:results} and \ref{sec:discussion}, respectively, followed by a summary in Section \ref{sec:summary}. Throughout this paper, we use the $\Lambda$CMD cosmology, with $H_{0} = 72.0$ $\rm km\ s^{-1}\ Mpc^{-1}$ and $\Omega_{m} = 0.3$.

\section{Data and Light-Curve Assembly}
{This section details the selection of H$\beta$ reverberation-mapped AGNs and the construction of optical and MIR light curves for the analysis.}
\label{sec:sample}

\subsection{Sample}
We collect AGNs with the H$\beta$ lag measurements from \cite{Wang2024}, who conducted a uniform $\rm H\beta$ RM analysis of 244 AGNs \textcolor{black}{($0.002 \lesssim z \lesssim 1.03$)}, comprising 212 AGNs with archival light curves (i.e., literature sample) and 32 AGNs from SAMP, employing the Interpolated Cross-Correlation Function \citep[ICCF,][]{Gaskell1987,Peterson1998} as their primary method. To ensure reliable $\rm H\beta$ lag measurements, \cite{Wang2024} implemented a rigorous quality assessment (QA). This QA process briefly involved several steps: (1) setting a threshold of $r_{\rm max} > 0.55$ for the maximum cross-correlation coefficient, (2) ensuring a $p$-value $<0.1$, (3) requiring a primary peak fraction $(f_{peak}) > 0.6 $ within the lag posterior distribution, followed by a visual inspection of the consistency between the $\rm H\beta$ and shifted continuum light curves (see their Section 3.5 for details). As a result, QA identified a sample of 157 AGNs with the best-quality lag measurements.

\textcolor{black}{We note that among the literature sample, some AGNs have multiple $\rm H\beta$ lag measurements derived from different archival light curves. For these cases, we retain only one lag measurement per target, selecting the one with the highest QA flags. When multiple measurements remain, we prioritize black hole mass estimates based on the velocity dispersion of the $\rm H\beta$ line from the rms spectra over those derived from the full width at half maximum (FWHM) of the mean spectra. Our final unique sample thus contains 182 AGNs, of which 126 have reliable $\rm H\beta$ lag measurements according to the QA criteria.} In the first part of this study, we use the full sample of $\rm H\beta$ reverberation-mapped AGNs to maximize the sample size for investigating the torus size-luminosity relation. In the second part, we restrict our analysis to the subset of AGNs with reliable $\rm H\beta$ lag measurements identified through the QA process to examine the relation between torus size and $\rm H\beta$ BLR size.

\subsection{Light-Curve Construction}
{Our light curve construction follows our previous work \cite[][]{Mandal2024}, and here, we briefly outline the process used in this analysis.}
\subsubsection{{Ground-based Optical Data}}
Optical light curves are constructed using photometric data from the Catalina Real-Time Transient Survey (CRTS; \citealt{Drake2009}), which used images from the Catalina Sky Survey (CSS) to detect optical transients. Initiated in 2004, CSS surveyed $\sim$33,000 $\rm deg^{2}$ of the northern sky with three telescopes; Catalina Schmidt Survey, Mount Lemmon Survey (MLS), and Siding Spring Survey (SSS), all using unfiltered CCDs. Data were calibrated to the Johnson $V$-band \citep{Drake2013}. The dataset includes time-series observations with a $\sim$2-3 week cadence for sources with $V \sim 20$ mag (Catalina Schmidt Survey and MLS, 2003-2016) and $V \sim 19$ mag (SSS, 2005-2013).  We retrieved photometric data from the public second data release and converted CRTS magnitudes to $V$-band fluxes using the method in \cite{Hovatta2014}.

Next, we gather photometric data from the All-Sky Automated Survey for Supernovae (ASAS-SN; \citealt{Shappee2014}), which detects bright transients using 24 telescopes across six global stations. Each station has four 14 cm telescopes with 2K $\times$ 2K CCDs, providing a $\rm 20\ deg^{2}$ field of view. Observations consist of dithered 90 s exposures, with photometric calibration via nearby APASS stars \citep{Henden2012}. ASAS-SN provides light curves for objects down to $V \sim$ 17 mag (2012-2018) and $g \sim$ 18 mag (2017-present). We collect $V$- and $g$-band flux light curves from the ASAS-SN Sky Patrol V2.0 \citep{Shappee2014, Hart2023}.

We incorporate photometric data from the Palomar Transient Factory (PTF; \citealt{Law2009}) and the Zwicky Transient Facility (ZTF; \citealt{Bellm2019}). From 2009 to 2014, PTF explored the optical transient sky with the Palomar 1.2 m Samuel Oschin Telescope, achieving $5 \sigma$ limiting magnitudes of 20.5 ($r$-band) and 21 ($g$-band) in 60 s exposures. In 2018, PTF transitioned to ZTF, which continues observations with an upgraded camera, widening the field-of-view and enhancing survey capabilities. ZTF offers a typical 3-day sampling interval and depths of $g \sim$ 20.8, $r \sim$ 20.6, and $i \sim$ 19.9 mag ($5 \sigma$ in 30 s). We incorporate PTF $r$-band data and ZTF $g$- and $r$-band light curves (DR23, 2018-2024).

Finally, for SAMP targets, in addition to the aforementioned data, we utilize light curves obtained between 2015 and 2021 with a cadence of 3-5 days \cite[][]{Woo2019,Woo2024}. Photometric monitoring was conducted using several telescopes, including the MDM 1.3 m and 2.4 m telescopes at Kitt Peak, Tucson, Arizona, USA; the Lemmon Optical Astronomy Observatory 1 m telescope on Mount Lemmon, Tucson, Arizona, USA; the Lick Observatory 1 m telescope at Mount Hamilton, California, USA; the Las Cumbres Observatory Global Telescope (LCOGT) network; and the Deokheung Optical Astronomy Observatory (DOAO) 1 m telescope. Observations were performed using B and V, or V and R, band filters to monitor continuum variability, depending on the redshift of each object \cite[][]{Woo2024}.

Since ZTF-$g$ and ASAS-SN-$g$ have overlapping coverage, we prioritize ZTF-$g$ data, which offers {$\sim$2-3 magnitudes greater depth, leading to a significantly higher signal-to-noise ratio.} When ZTF-$g$ data is unavailable, we use ASAS-SN-$g$ as complementary data. All magnitudes are converted to flux, and the flux units are standardized to mJy.

\subsubsection{Photometric Calibration of Optical Light Curves}
To finalize the construction of optical light curves, we perform intercalibration to correct systematic offsets between photometric data from different telescopes at different times, potentially caused by varying weather conditions, filter responses, detector efficiencies, etc. We utilize PyCALI \citep[][]{Li2014}, a Bayesian Python package, which employs a damped random walk \citep[DRW,][]{Kelly2009,MacLeod2010} model to characterize AGN variability and optimize scaling factors. We designate ASAS-SN $V$-band data as the reference light curve when available due to its high cadence and its placement at the core of the overall light curve; otherwise, we use ZTF $g$-band. Given that optical interband continuum lags, estimated to scale to a few days \citep[e.g.,][]{Jiang2017,Guo2022} and being much smaller than the MIR timescales, they are therefore disregarded in our intercalibration process. To further enhance the reliability and consistency of the calibrated light curves, a systematic error term was included to account for flux measurement uncertainties.

\subsubsection{{WISE MIR data}}
\label{sec:wise_lc}
WISE performed an all-sky survey with images at 3.4, 4.6, 12, and 22 $\rm \mu m$ (W1, W2, W3, and W4) over ten months in 2010 using a 40 cm diameter IR telescope. After completing its primary mission, it was reactivated in 2013 as the Near-Earth Object Wide-field Infrared Survey Explorer \citep[NEOWISE,][]{Mainzer2011}, focusing on detecting near-Earth objects. We construct MIR light curves incorporating the full mission data set by collecting photometric data in W1 and W2 until August 2024. To ensure reliability, we select database entries meeting specific quality criteria as suggested in {\url{https://wise2.ipac.caltech.edu/docs/release/neowise/expsup/sec2_3.html}}. Briefly, we require high overall frame quality (\texttt{qual\_frame > 0}), best frame image quality (\texttt{qi\_fact = 1}), sufficient distance from the South Atlantic Anomaly (\texttt{saa\_sep $\geq$ 5}), no moon masking (\texttt{moon\_masked = \textquotesingle 00\textquotesingle}), and no confusion or contamination flags (\texttt{cc\_flags = \textquotesingle 00\textquotesingle}). WISE data exhibit 180-day gaps between cycles. We bin W1 and W2 data within 180-day windows to produce evenly sampled light curves and calculate the mean flux. WISE magnitudes, derived from single-exposure images using profile-fitting photometry, are converted to flux using the zero-point flux densities from \cite{Jarrett2013}.

The MIR fluxes in the W1 and W2 bands contain emission from both the dusty torus and the accretion disk \citep[e.g.,][]{Kishimoto2008,Lira2011}. The accretion disk, which varies on shorter timescales, introduces a prompt, undelayed component to the MIR light curves, biasing the measured lag toward shorter values. \cite{Mandal2024} showed that correcting for this contamination increases torus size estimates, especially at higher AGN luminosities. This effect is stronger in W1 than in W2 and shows no significant dependence on the Eddington ratio. We correct the accretion-disk contamination following \cite{Mandal2024} to recover the intrinsic torus lag. \textcolor{black}{We also note that, in our moderate-to-high-luminosity type 1 AGNs, hot dust emission from the AGN torus dominates over host galaxy stellar emission, resulting in the stellar contribution to the W1 and W2 bands being negligible.} We estimate the disk component for each W1 and W2 epoch by averaging optical flux from observations within $\pm$ 5 days. If no such data were available, we used optical fluxes predicted from a DRW-modeled light curve generated with PyCALI. We subtracted the estimated disk contribution from the MIR light curves to isolate the torus emission.

\section{Data Analysis}
\label{sec:data_analysis}
After constructing our optical and MIR light curves, we explore the time delays between these bands to determine dust lag and torus size. Given the presence of $\rm H\beta$ lag measurements of our sample, which trace variability in the optical continuum from the central accretion disk, corresponding variations in the MIR emission are expected. This investigation utilizes the ICCF as its primary method, with the Multiple and Inhomogeneous Component Analysis (MICA) serving as a secondary approach detailed in Sections \ref{sec:sub1}-\ref{sec:sub4}.
\subsection{{Dust Lag Measurements}}
\label{sec:sub1}
As the primary method for measuring dust lags, we adopt the widely used ICCF technique, implemented via the Python package \texttt{PyCCF} \citep[][]{Peterson1998,Sun2018}. This method calculates the cross-correlation function (CCF) between two light curves by linearly interpolating one onto the time grid of the other and shifting it across a range of time delays. To reduce interpolation bias, the procedure is repeated with the interpolation direction reversed, and the final CCF is the average of the two. For each time delay, the Pearson correlation coefficient ($r$) is used to quantify the similarity between the light curves. The time lag is estimated by identifying either the peak of the CCF or the centroid, a weighted average of the CCF above 80\% of the peak value ($r_{\rm max}$). Although peak and centroid lags show minor differences \citep[][]{Peterson2004}, the centroid lag is preferred for higher accuracy. The peak lag value is sensitive to noise, causing greater uncertainties \citep[][]{White1994,Peterson1998,Peterson2004}. Thus, the centroid lag is a more reliable time delay estimate \citep[see][for more details]{Gaskell2025}. 

For our analysis, we search lag values within a window of [--1000, 2000] days. We estimate the uncertainties in lag determination through flux randomization (FR) and random subset selection (RSS) methods \citep[e.g.,][]{Peterson1998}. Briefly, in FR, flux values are randomly perturbed within their respective uncertainties, while in RSS, subsets of observed epochs are randomly selected, and the lag is recalculated. We perform 5000 FR/RSS realizations to construct a cross-correlation centroid distribution (CCCD), adopting the median as the best estimate of the lag and the 15.87th and 84.13th percentiles as the uncertainty bounds.  {For some targets, secondary peaks appear at larger lags, likely caused by aliasing effects due to variability amplitude, seasonal gaps, or cadence limitations.} For example, \citet{Mandal2024} found that in some cases, these secondary peaks are associated with the autocorrelation structure in the optical light curves. Therefore, such peaks are generally not interpreted as true physical delays. In our sample, only one target (RBS1303) exhibits a secondary peak of this kind. We focus on the primary peak to determine the time lag for this case.

\begin{figure*}[htbp]
  \centering
  \includegraphics[scale=0.58]{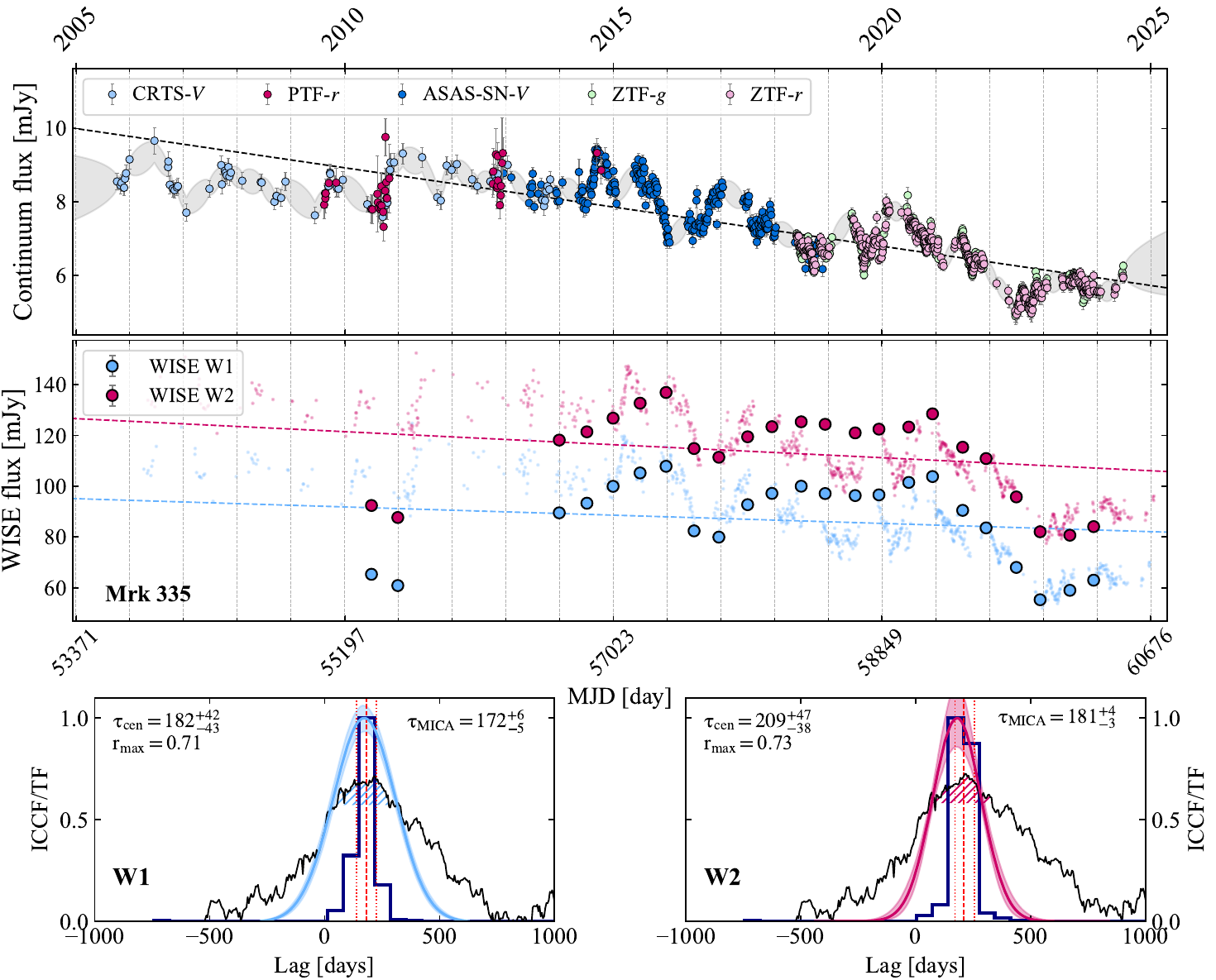}
  \caption{Top: Lag analysis result for Mrk 335, shown as an example of a detrended target. The top panel shows intercalibrated optical light curves, with the gray-shaded regions representing the DRW model from PyCALI. The middle panel presents accretion-disk-corrected MIR light curves in the WISE W1 and W2 bands. The blue and red points represent the scaled optical light curves shifted by the $\tau_{cent}$ from ICCF relative to the W1 and W2 light curves, respectively. The dashed lines show the first-order polynomial fit on light curves. The bottom-left panel shows the ICCF (black curve) and CCCD (blue histogram) for W1. The hatched region below the ICCF curve indicates where $r \geq 0.8$ r$_{\rm max}$. The best-fit transfer function derived from the MICA analysis is shown in blue, with the shaded region indicating uncertainties. The bottom-right panel shows W2 data, mirroring the left panel.}
  \label{fig:example1}
\end{figure*}
  
\begin{figure*}[htbp]
  \centering
  \includegraphics[scale=0.58]{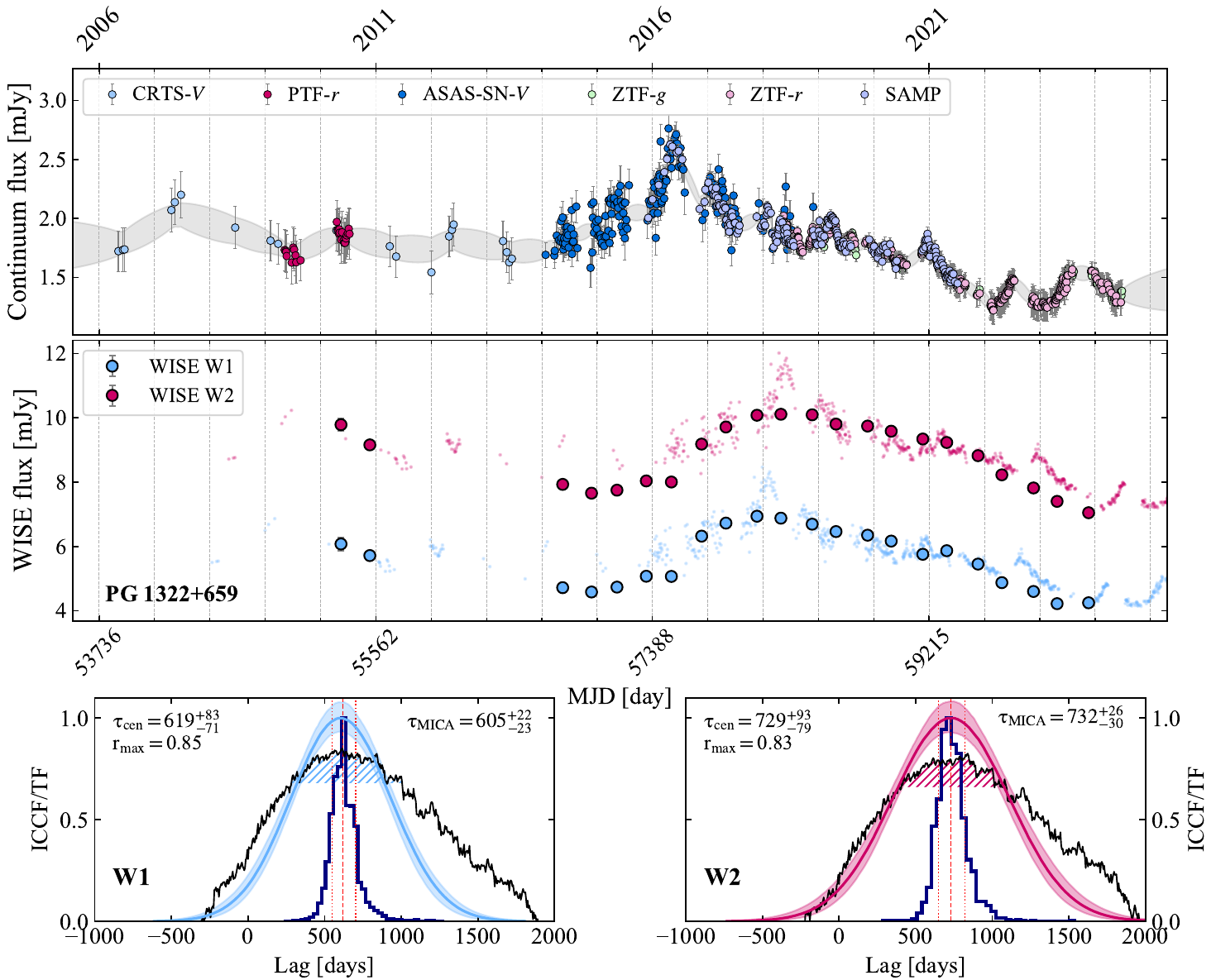}
  \caption{The same as Figure \ref{fig:example1} showing lag analysis for PG 1322+659, an example from the SAMP sample.}
  \label{fig:example2}
\end{figure*}

We also assess dust lag estimation through the MICA\footnote{\url{https://mica2.readthedocs.io/en/latest/}} \citep[][]{Li2016}, a nonparametric RM method that estimates time lags \citep[e.g.,][]{Chelouche2019,Chen2023,Li2024} by modeling the relationship between driving and responding light curves through a transfer function. Although MICA supports a range of functional forms, including top hat, gamma, and exponential profiles, we adopt the Gaussian model. Hence, the transfer function is represented as a sum of relatively displaced Gaussian components, where the center of each Gaussian corresponds to a characteristic time lag. We further simplify the approach by using a single Gaussian component. The posterior distribution of the Gaussian center is sampled using a Markov Chain Monte Carlo (MCMC) technique, yielding 5000 realizations of the transfer function. The median of this distribution is taken as the best estimate of the time lag, while the 15.87th and 84.13th percentiles define the associated uncertainties. Figures \ref{fig:example1} and \ref{fig:example2} present two examples of lag measurement results.

Long-term trends in light curves can bias lag measurements by masking shorter timescale variations associated with the true delay \citep[e.g.,][]{Zhang2019,Chen2023}. To mitigate this, we fit and subtract a first-order polynomial from the continuum and MIR light curves prior to measuring the lag. We identify such trends in five objects and apply this detrending procedure accordingly (see Figure \ref{fig:example1}). 

\subsection{{Quality Assessment}}
\label{sec:sub2}
Following our previous studies \citep[][]{Woo2024, Mandal2024,Wang2024}, we apply three key selection criteria to ensure robust and reliable time lag measurements from the ICCF. First, we require a strong correlation between optical and MIR light curves, setting a conservative threshold of $r_{\text{max}} > 0.6$. Although previous studies have used lower limits \citep[e.g., $r_{\text{max}} > 0.4$,][]{Chen2023}, our stricter cutoff helps reduce spurious or misleading correlations. However, a high $r_{\text{max}}$ does not always imply a genuine physical delay. Factors such as autocorrelation in sparsely sampled or smoothly varying light curves can produce {aliasing peaks} in the CCF, leading to false detections \citep[e.g.,][]{Welsh1999}. These issues highlight the need for additional criteria beyond $r_{\text{max}}$ to assess lag reliability.

As a second criterion, we evaluate the likelihood that the observed correlation occurs by chance, following previous multiwavelength AGN variability studies \citep[e.g.,][]{Uttley2003,Arevalo2008,Chatterjee2008,Cameron2012,Yu2023,Woo2024,Mandal2024}. This significance depends on $r_{\text{max}}$, as well as the strength and coherence of the light-curve variability. While prominent, sustained features aid detection; sporadic flares can produce spuriously high $r_{\text{max}}$. To assess significance, we follow \citet{Wang2024} by generating 5000 mock MIR light curves with randomized fluxes but preserved sampling and trends. These are cross-correlated with real optical data, and we define significant lags as those with $p(r_{\text{max}}) \leq 0.1$.

\begin{figure}[ht!]
\centering
\includegraphics[width=0.475\textwidth]{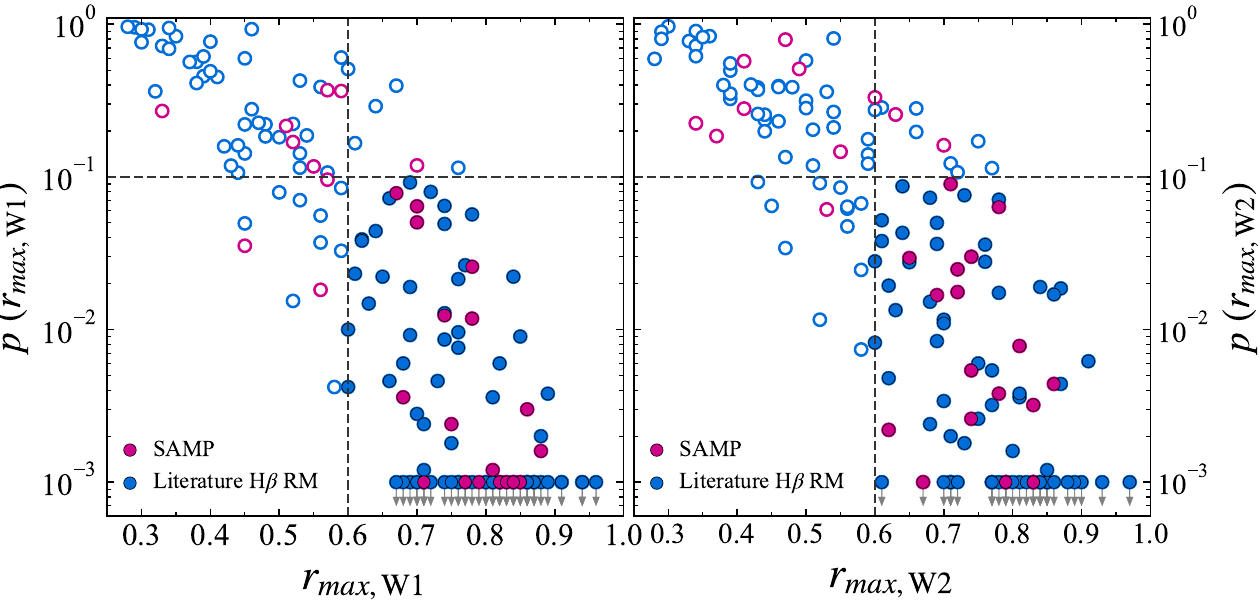}
\caption{Distribution of $p(r_{\rm max})$ versus $r_{\rm max}$ for W1 (left) and W2 (right). Blue circles represent the H$\beta$ RM literature sample, and red circles represent SAMP AGNs. The black dashed lines indicate the selection criteria of $p(r_{\rm max}) = 0.1$ versus $r_{\rm max} = 0.6$. Targets that do not meet these criteria are presented as open circles.}
\label{fig:rmax_pval}
\end{figure}

Finally, we visually inspect the ICCF results using the calculated time lags to confirm the reliability of our findings. During this step, we look for a clear, prominent peak in the correlation, check for any misleading secondary peaks, and ensure that the results align with the expected behavior of the light curves. This visual inspection validates the significance of the time delays and adds confidence to our overall analysis. Applying these criteria, we identify 92 AGNs with reliable W1 lag measurements and 85 AGNs with reliable W2 lag measurements. Figure \ref{fig:rmax_pval} presents the distribution of $p(r_{\rm max})$ versus $r_{\rm max}$ for these measurements.

\subsection{{Comparison of Time Lags from ICCF and MICA}}
\label{sec:sub3}
\begin{figure*}[ht!]
\centering
\includegraphics[width=0.92\textwidth]{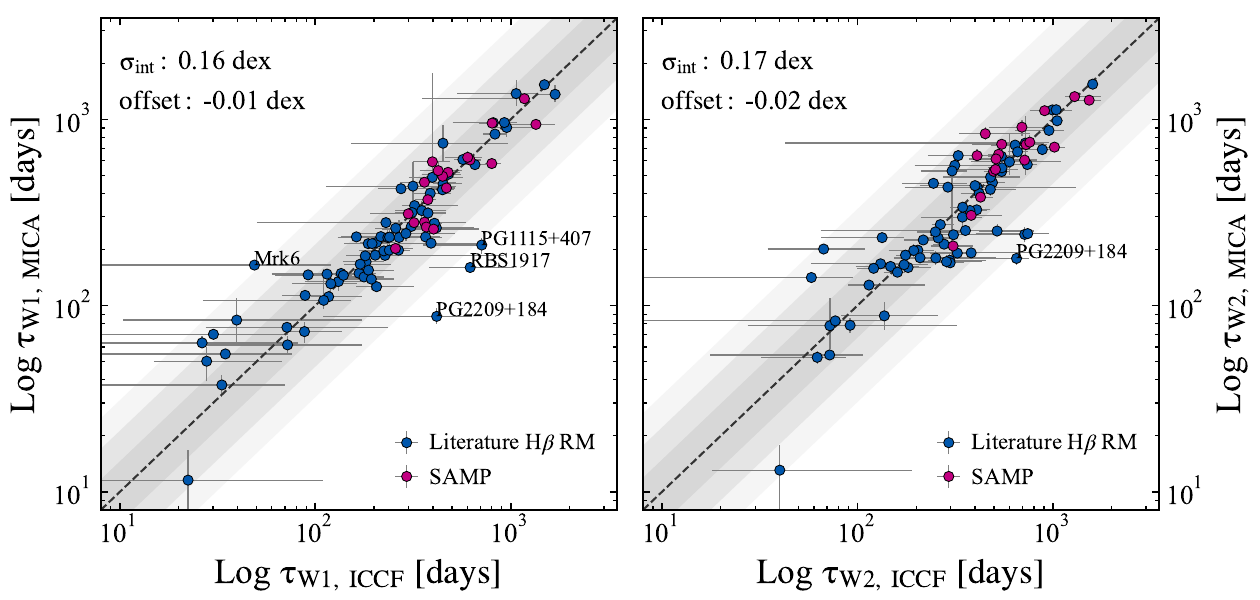}
\caption{Comparison of lag measurements obtained using ICCF and MICA in the observed frame for WISE W1 (left) and W2 (right). Colors follow the same convention as in Figure \ref{fig:rmax_pval}. 
The shaded areas show  $\pm1\sigma_{\rm int}$, $\pm2\sigma_{\rm int}$, and $\pm3\sigma_{\rm int}$  intrinsic scatter around one-to-one line, illustrating the agreement between the two methods.}
\label{fig:iccf_mica}
\end{figure*}

To further assess the reliability of our results, we compare the ICCF and MICA lag measurements and show consistency between them with small scatters and negligible offsets (i.e., $\sigma_{\rm int} = 0.16$ dex and offset = $-0.01$ dex for W1 and $\sigma_{\rm int} = 0.17$ dex and offset = $-0.02$ dex for W2) in Figure~\ref{fig:iccf_mica}, reinforcing the robustness of our selected time lags. Compared to the scatter of $\sim$2.4 dex (corresponding to a standard deviation of $\sim$276 days) dispersion between ICCF and MICA measurements reported by \citet{Chen2023}, our results show better agreement, with standard deviations of $\sim$2.1 dex ($\sim$125 days) for W1 and $\sim$2.2 dex ($\sim$150 days) for W2. 

Our better consistency likely arises from the enhanced quality of our light curves, driven by a longer, more densely sampled baseline and more reliable lag measurements from stricter selection criteria. For example, \citet{Chen2023} reported significant discrepancies in the lag measurements of Mrk 335, with MICA lags of $\sim$$169^{+9.6}_{-12}$ and $195^{+16.1}_{-11.2}$  days for W1 and W2, respectively, compared to ICCF lags of $721 ^{+521}_{-318}$ and $725^{+359}_{-468}$ days. In contrast, as shown in Figure \ref{fig:example1}, we find good agreement between the two methods for this target. Nevertheless, we identify four outliers (PG2209+184, PG1115+407, Mrk 6, and RBS 1917) that deviate beyond the 3$\sigma_{\rm int}$ dispersion range in the W1 band. The lag values of our sample are presented in Table \ref{tab:final_lags}.

\subsection{{Redshift Correction of Dust Time Lags}}
\label{sec:sub4}
The measured time lags are influenced by both cosmological time dilation and the redshift of the wavelength for given photometric bands. 
These lags reflect the light-travel time to dust located at varying distances and temperatures within the obscuring torus structure, depending on the redshift of the target. As temperature increases with decreasing distance from the central engine, shorter rest-frame wavelength ($\lambda$) of higher-z AGNs corresponds to smaller rest-frame lags. Thus, it is necessary to correct for the redshift effect on the measured time lag for a given band. 

The dust sublimation radius ($R_{\mathrm{sub}}$) is theoretically related to the AGN luminosity ($L_{\mathrm{AGN}}$) and the dust sublimation temperature ($T_{\mathrm{sub}}$) through the relation $R_{\mathrm{sub}} \propto \left( {L_{\mathrm{AGN}}}/{10^{45}\mathrm{erg,s^{-1}}} \right)^{1/2} \left({1500\mathrm{K}}/{T_{\mathrm{sub}}} \right)^{\gamma}~\mathrm{pc}$ \citep[][]{Barvainis1987,Nenkova2008}, which implies $T_{\mathrm{sub}} \propto R_{\mathrm{sub}}^{-1/\gamma}$. Since $R_{\mathrm{sub}} \propto \tau / c$ and dust at temperature $T_{\mathrm{sub}}$ emits primarily at $\lambda \sim hc / 4kT$, the rest-frame lag is expected to follow a power-law relation with wavelength as $\tau \propto \lambda^{\gamma}$. The exponent $\gamma$ is theoretically estimated to be 2.8 \citep[][]{Barvainis1987} or 2.6 \citep[][]{Nenkova2008}.
At higher redshifts, observations at a given wavelength probe progressively shorter rest-frame values. Consequently, high-redshift AGNs tend to show intrinsically shorter rest-frame dust lags for a given luminosity compared to their low-redshift counterparts. To ensure consistent comparisons of dust lag measurements across different redshifts, observed time lags are corrected by a factor of $(1+z)^{\gamma - 1}$, providing rest-frame values normalized to a common wavelength for consistent cross-redshift comparisons \citep[e.g.,][]{Oknyanskij2001, Minezaki2019, Mandal2024}.

In this work, we estimate the value of $\gamma$ for our sample using the relation $\gamma = \log(\tau_{\text{W2}} / \tau_{\text{W1}}) / \log(\lambda_{\text{W2}} / \lambda_{\text{W1}})$. We find the median ratios of $\tau_{\text{W2}} / \tau_{\text{W1}} \sim 1.25 \pm 0.07$ (corresponding to the  $\gamma \sim 0.74$), which is in agreement with previously reported values of 1.22 \citep{Mandal2024}, 1.26 \citep{Chen2023}, and 1.15 \citep{Lyu2019}. The significant discrepancy between the empirical and theoretical values of the factor $\gamma$ may be partly explained by the observed shallower slope of the dust lag-luminosity relation. This shallower slope implies that the size of the dust-emitting region scales less strongly with luminosity than anticipated, consequently weakening the observed dependence of lag on wavelength. We apply the correction factor of 
$(1+z)^{-0.26}$ to the observed-frame time lags to obtain the rest-frame lags.

\section{Results}
\label{sec:results}
\subsection{Dependence of Torus Size on AGN Luminosity}
\label{sec:size-opt}

We investigate the correlation between dust torus size and AGN optical luminosity using rest-frame time lags by fitting a linear relation,
\begin{equation}
\log (R_{\mathrm{tor}}/\text{1t-day}) = \alpha + \beta \log \left(L_{\rm AGN}/\mathrm{10^{44}\ erg\, s^{-1}} \right)
\label{equ:fit}
\end{equation}
where $R_{\mathrm{tor}}$ represents the dust lag in light-days (ld) and $L_{\rm AGN}$ is AGN luminosity. We perform the regression using the {\tt linmix} package \citep{Kelly2007}, which accounts for measurement uncertainties in both variables and the intrinsic scatter in the correlation. We use the monochromatic luminosity at 5100 \AA\ as a proxy for AGN luminosity. \textcolor{black}{While our sample is composed of moderate-to-high-luminosity AGNs, some of them show signatures of stellar contamination. Thus, we use host-corrected AGN continuum luminosities at 5100 \AA\ \citep[][]{ Woo2024,Wang2024}.} We first allow the slope to vary freely and then repeat the fit with the slope fixed to 0.5. We obtain the best-fit parameters for W1 and W2 as follows:
\begin{align}
\log \left(R_{\mathrm{tor,W1}}/\text{1t-day} \right)
&= 2.45 \pm 0.02\ + \notag \\
&\quad 0.35\pm 0.03 \log \left(L_{5100}/\mathrm{10^{44}\ erg\, s^{-1}} \right)
\end{align}
\begin{align}
\log \left(R_{\mathrm{tor,W2}}/\text{1t-day} \right)
&= 2.59 \pm 0.02\ + \notag \\
&\quad 0.33 \pm 0.03 \log \left( L_{5100}/10^{44}\ \mathrm{erg\, s^{-1}} \right)
\end{align}

\begin{figure*}[ht!]
\centering
\includegraphics[width=0.92\textwidth]{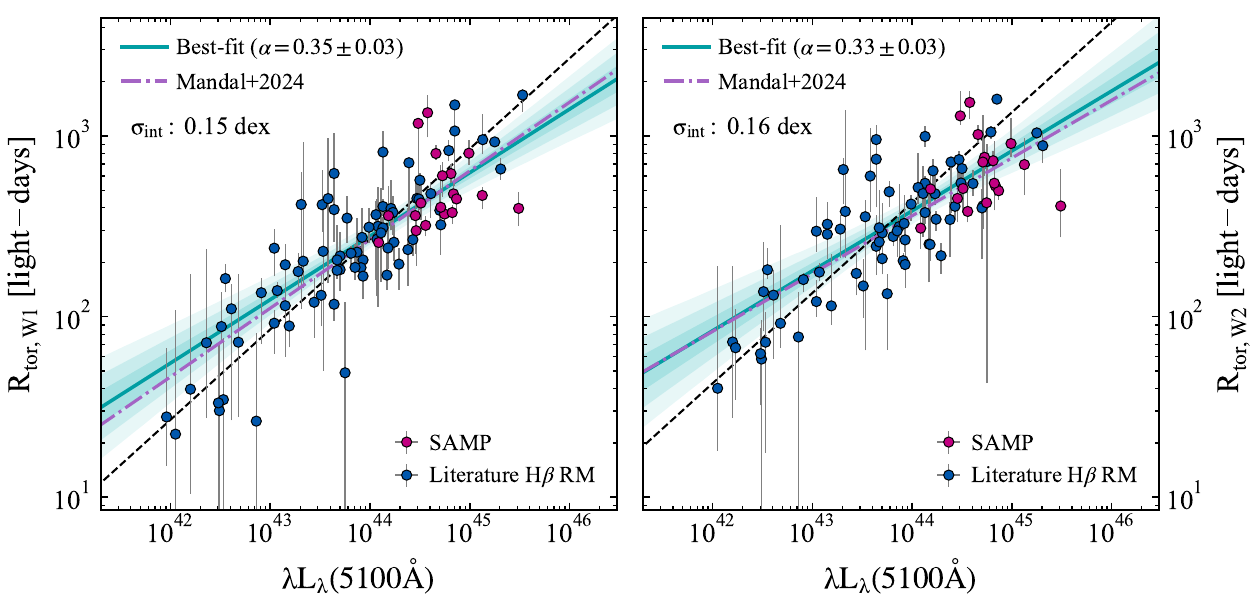}
\caption{Size of torus ($R_{\rm tor}$) as a function of 5100 \AA\ luminosity for W1 (left) and W2 (right). The green solid line represents our best-fit regression with a free slope. Shaded regions represent the $1\sigma$, $2\sigma$, and $3\sigma$ confidence intervals around the best-fit regression line, derived from the percentiles of model predictions sampled from the posterior distributions of slope and intercept (hereafter this method is applied in all related figures). The black dashed lines represent the best-fit model with a fixed slope of 0.5, while the purple lines show the regression results from \cite{Mandal2024}.}
\label{fig:size_lum}
\end{figure*}

We obtain a clear sub-linear torus size-luminosity relation, as shown in Figure \ref{fig:size_lum}. The best-fit slopes are $0.35\pm 0.03$ and $0.33\pm 0.03$ with intrinsic scatters of 0.15 dex and 0.16 dex, for W1 and W2, respectively.  The observed slopes are shallower than the theoretical prediction. This discrepancy implies that additional physical effects may be important and are discussed further. \textcolor{black}{For both W1 and W2 bands, the covariance between $\alpha$ and $\beta$ is close to zero, and the correlation coefficients are $\sim -0.15$, consistent with the choice of a pivot luminosity near the median.}

Our results are consistent with previous studies. Notably, \citet{Mandal2024} reported $\beta = 0.38 \pm 0.02$ for W1 and $\beta = 0.32 \pm 0.02$ for W2, values derived from a larger sample, while their analysis is consistent with the current study since both studies utilized the ICCF method for lag measurements and adopted comparable criteria for identifying reliable lags. 
\citet{Chen2023} also reported comparable slopes of $\beta = 0.365 \pm 0.029$ for W1 and $\beta = 0.372 \pm 0.027$ for W2, based on a sample that largely overlaps with ours, as it consists of AGNs with H$\beta$ RM measurements. However, they used MICA-based lag measurements and did not correct MIR light curves for accretion-disk contamination. \citet{Mandal2024} demonstrated that correcting for accretion-disk contamination leads to systematically larger torus sizes and steeper lag-luminosity slopes, particularly in the W1 band where the accretion-disk contamination is more significant. These results underscore the importance of applying accretion disk-correction when interpreting MIR lags, especially at shorter wavelengths. For a fixed slope of 0.5, we obtain intercept values of $\alpha = 2.43 \pm 0.01$ for W1 and $\alpha = 2.63 \pm 0.01$ for W2. The results of our regression analysis are summarized in Table \ref{tab:regression_parameters}.

{{In comparison, \cite{Gravity2024} derived dust size-luminosity relations of $R_{\rm tor} \propto L_{\rm 5100}^{0.39}$ from NIR interferometric measurements and $R_{\rm tor} \propto L_{\rm 5100}^{0.37}$ from RM data, both based on the monochromatic luminosity at 5100 \AA. By applying a nonlinear bolometric correction to estimate the bolometric luminosity, they found that the slopes increased to 0.46 and 0.44 for the interferometric and RM samples, respectively \cite[see also][]{Lyu2019}.} The nonlinear bolometric corrections may provide a more accurate conversion from monochromatic to bolometric luminosities and tend to yield dust size-luminosity relations that better align with theoretical expectations \citep[][]{Lyu2019,Gravity2024}. As a result, \cite{Gravity2024} suggested that comparing the slope of the size-luminosity relation derived from monochromatic luminosities (or those using a simple linear correction) to the canonical value of 0.5 may lead to misinterpretation. This is largely due to the fact that the connection between monochromatic luminosity and dust heating depends on the shape of the SED. To test this scenario, following \cite{Lyu2019}, we convert the 5100\AA\ luminosity to the bolometric luminosity using the nonlinear bolometric correction from \cite{Runnoe2012}:
\begin{equation}
\log (L_{\mathrm{AGN, bol}}/\mathrm{ erg\, s^{-1}}) = 4.89 + 0.91 \log \left(L_{5100}/\mathrm{erg\, s^{-1}} \right)
\end{equation}
where $L_{\mathrm{AGN, bol}}$ is AGN bolometric luminosity. We repeat the regression using bolometric luminosity, and the slope of the torus size-luminosity relation slightly increases to $0.39 \pm 0.03$ in W1 and $0.35 \pm 0.04$ in W2. {If the torus size scales as $R_{\rm tor} \propto L^{0.5}$, our adopted bolometric correction naturally modifies slope compared to using monochromatic luminosity. Therefore, our conclusion regarding the torus size-bolometric luminosity slope is contingent on the validity and form of the assumed bolometric correction.}

For a consistency check, we combine our sample with that of \citet{Mandal2024} to perform regression, after excluding 12 duplicated sources from the \citet{Mandal2024} sample. We obtain the best-fit slope of $\beta = 0.36 \pm 0.01$ for the W1 band and $\beta = 0.31 \pm 0.01$ for W2, which are consistent with the best fit based on our H$\beta$ RM sample (see Figure~\ref{fig:size_lum_all}).

\begin{figure*}[ht!]
\centering
\includegraphics[width=0.92\textwidth]{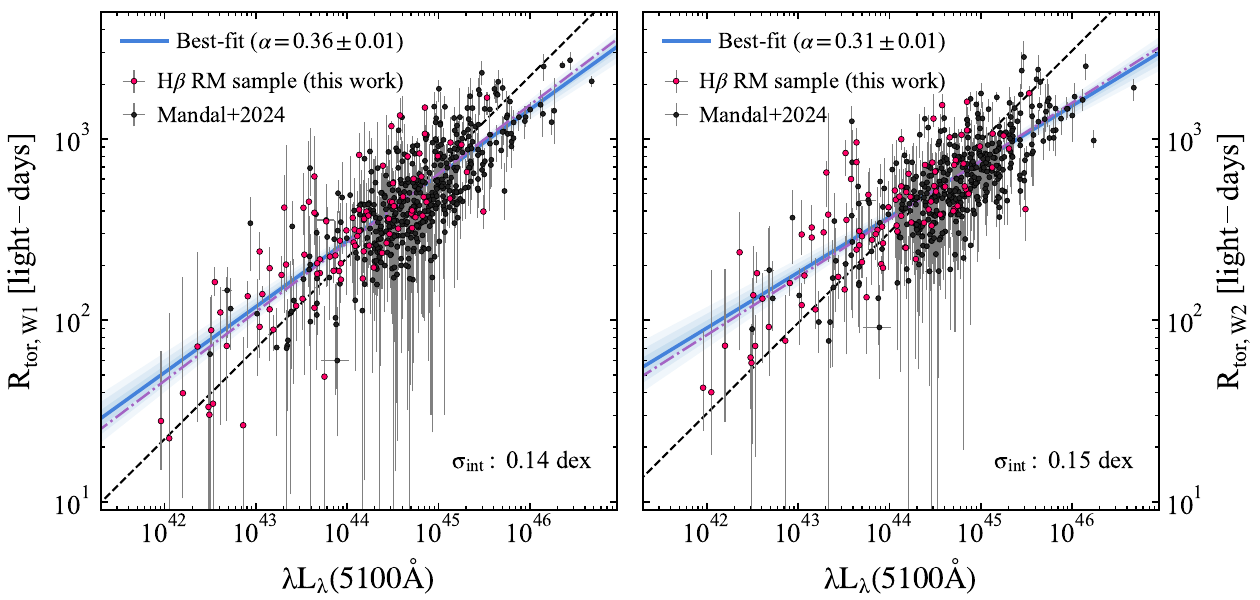}
\caption{Size of torus ($R_{\rm tor}$) as a function of 5100 \AA\ luminosity for W1 (left) and W2 (right) for merged sample. The black circles illustrate the sample from \citet{Mandal2024} while red circles indicate the sample of this study. The blue solid line shows our best fit with a free slope, and the shaded regions represent the $1\sigma-3\sigma$ confidence intervals around the best fit. The purple lines show the regression results from \citet{Mandal2024}, while the black dashed line marks the fit with a fixed slope of 0.5.}
\label{fig:size_lum_all}
\end{figure*}

\textcolor{black}{\subsubsection{Is Limited Time Resolution Driving a Shallower Slope?}}
\textcolor{black}{Given the $\sim$180-day cadence of WISE observations, it can be challenging to measure the lags of low-luminosity AGNs when the expected lags are shorter than the observational time resolution. On the other hand, variability amplitude or structures in the light curve are important for measuring lags regardless of the time resolution. To assess potential biases arising from undetected lags due to the time resolution limitation, we compare the luminosity distributions of variable AGNs with and without measured lags. 
We identify variable AGNs using the fractional rms variability amplitude ($\rm F_{var}$). We apply a threshold of $\rm F_{var} > 0.05$ to select a subset of AGNs exhibiting significant variability, which is necessary for reliable lag detection. Figure \ref{fig:lum-variables} illustrates the 5100 \AA\ luminosity distribution for both samples. We use the two-sample Kolmogorov-Smirnov (KS) test to compare the luminosity distributions, obtaining a statistic of 0.13 and a $p$-value of 0.49. This indicates no statistically significant difference between the two, both of which have median values of $\log (L_{5100}/\rm erg\ s^{-1}) \sim 44.1$, suggesting that the absence of lag measurements due to time resolution is not strongly biased toward low-luminosity targets. Moreover, an AGN with luminosity $\log (L_{5100}/\rm erg\ s^{-1}) \lesssim  43.7$ is expected to have a lag $ \lesssim 180$ days, corresponding to the WISE cadence (dashed line in Figure~\ref{fig:lum-variables}). Among our variable targets, 43 AGNs have luminosities below this threshold. Of these, 27 AGNs ($\sim$ 65\%) have reliable lag measurements, while the remaining 16 do not have measured lags. These results suggest that the limited time resolution is not the main factor for the failure of lag measurements. For many high-luminosity AGNs, the lag was not measured presumably due to the lack of strong patterns in the light curves.}

\begin{figure}[htbp]
  \centering
  \includegraphics[scale=0.58]{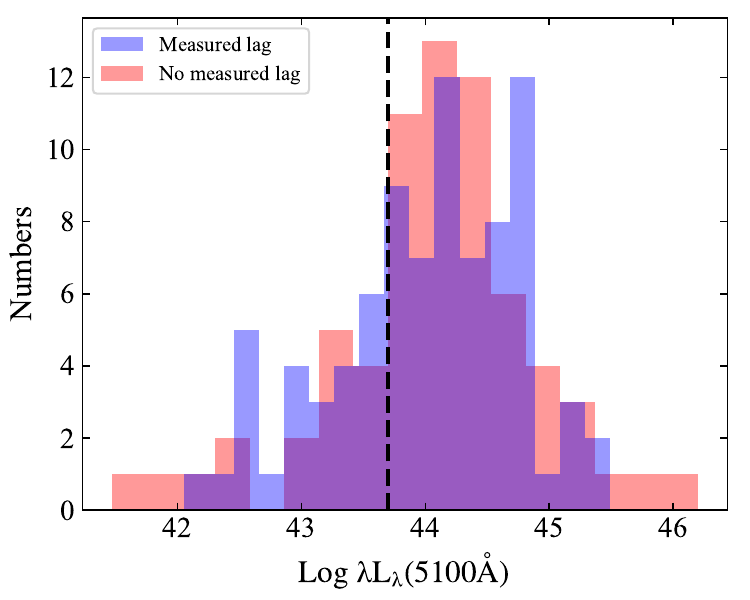}
  \caption{The distribution of 5100\AA\ continuum luminosity for variable AGNs with (blue) and without (red) measured lags. The black dashed line indicates the luminosity at which a lag of $\sim$180 days is expected, corresponding to the WISE cadence.}
  \label{fig:lum-variables}
\end{figure}

\textcolor{black}{Furthermore, we conduct Monte Carlo simulations for $\sim$ 50\% of AGNs without measured lags, selected to cover the full luminosity range, to assess the detectability of lags in these sources. We generate a broad lag distribution of 1000 values centered on the expected lag from the dust size-luminosity relation. From this distribution, we select representative lags corresponding to the 1\%, 15\%, 50\%, 85\%, and 99\% quantiles. If large lags are not already sampled, we additionally include longer lags up to $\sim$ 200 days to ensure that possible long lags are considered.}
\textcolor{black}{We simulate MIR light curves by sampling DRW realizations of the observed optical light curves by randomizing the Gaussian noise, such that the resulting variability amplitudes reflect the observed variability of each target.} The light curves are then time-shifted by input lags to represent the expected delay between optical and MIR emission. For each input lag, we generate 50 simulated MIR light curves using the actual WISE observation times and noise characteristics to reproduce realistic variability. Lags are then measured with the ICCF method. We define the true lag detection probability as the fraction of trials where the recovered lag is within 30\% of the input value.

From our simulations, we generally find that for AGNs with luminosity below $\log (L_{5100}/\rm erg\ s^{-1}) \sim 43.7$, where the expected lag is shorter than the WISE cadence, the probability of detecting a true lag increases with increasing input lag (Figure \ref{fig:simulations}, left panel). However, a subset of targets consistently shows low true lag detection probabilities across all input lags, suggesting that factors beyond cadence, such as low variability amplitude or data quality, may also play a role. For higher-luminosity AGNs, there is a tendency for detection probabilities to remain low across much of the input lag range (Figure \ref{fig:simulations}, right panel). This is likely due to limited variability amplitude, which reduces the ability to reliably detect lags even when the time resolution is not a limiting factor.

\begin{figure*}
\centering
\setkeys{Gin}{width=0.4\linewidth}
\includegraphics{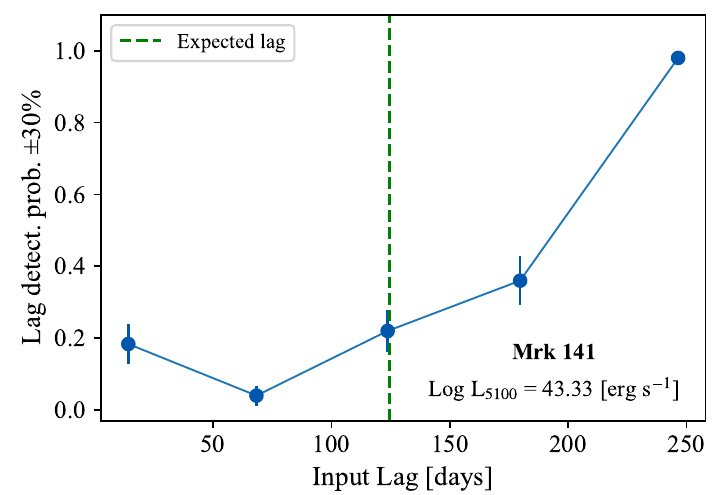}
\includegraphics{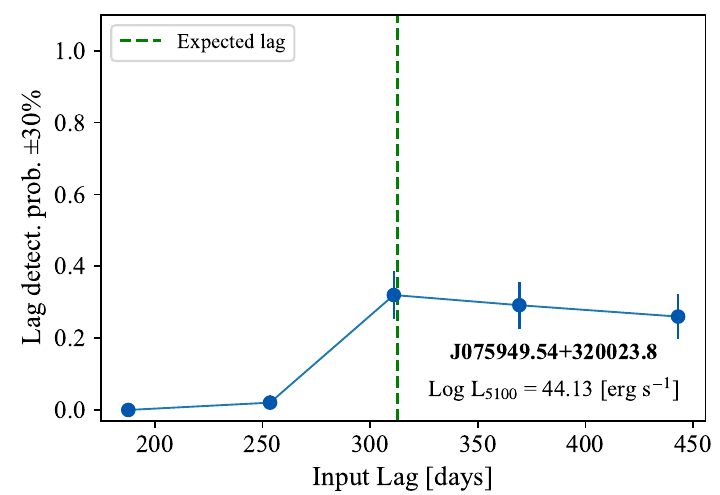}
\caption{Examples of \textcolor{black}{true} lag detection probability as a function of input lag from Monte Carlo simulations for AGNs without measured lags. The green dashed line indicates the expected lag from the dust size-luminosity relation assuming a slope of 0.5.}
\label{fig:simulations}
\end{figure*}

While these trends reflect general behaviors in our sample, the detection probability can vary significantly from target to target. Overall, while the WISE time resolution may lead to missing a fraction of short lags at low luminosities, short lags are still recovered for the majority of low-luminosity AGNs in our sample. Thus, there is no strong evidence that this bias is responsible for the observed shallower slope.}

\begin{figure*}[ht!]
\centering
\includegraphics[width=\textwidth]{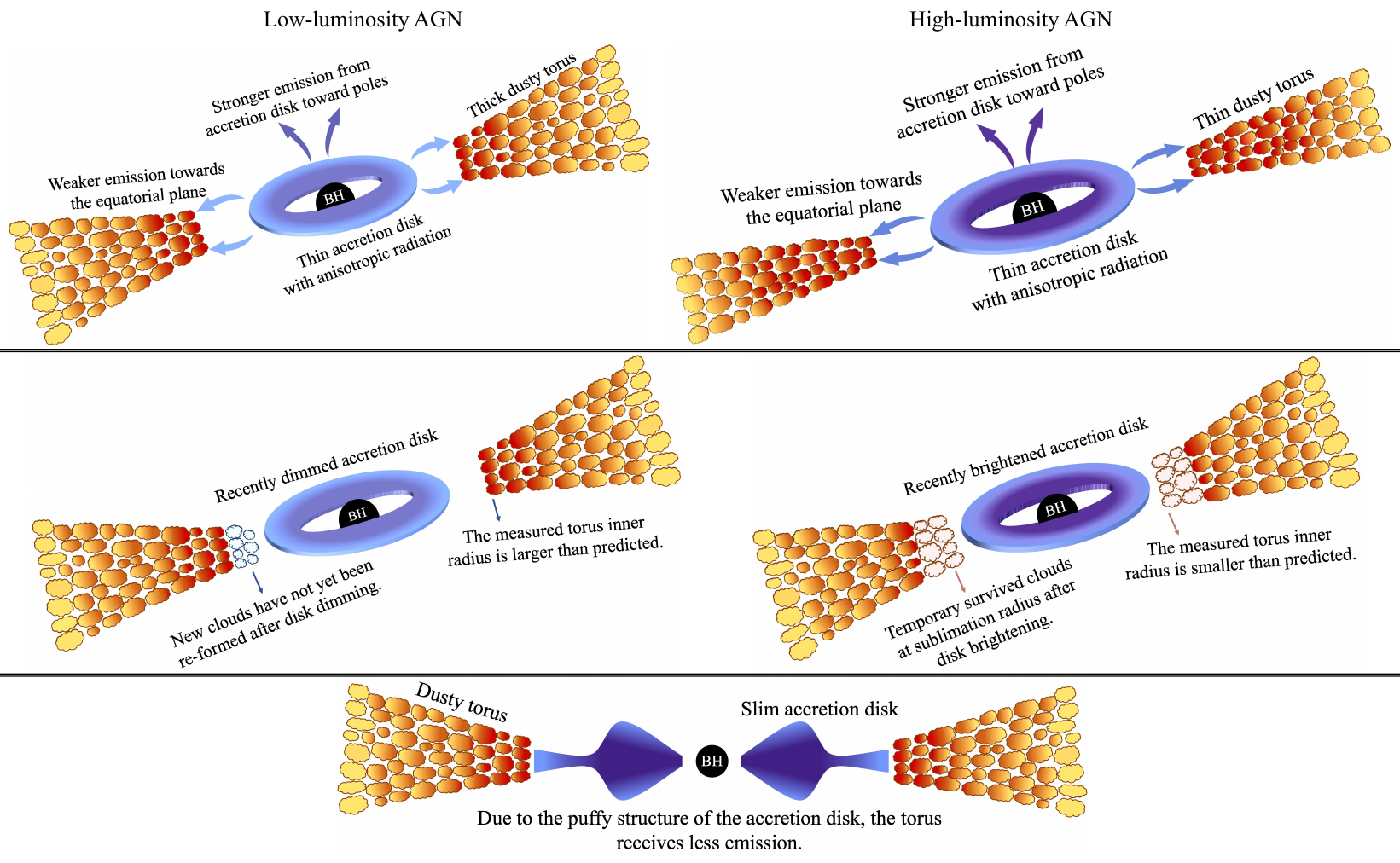}
\caption{Schematic illustration of possible scenarios that may explain the shallower-than-expected slope of the torus size-luminosity relation. The top row shows the effects of anisotropic disk emission and torus geometry in low-luminosity (left) and high-luminosity (right) AGNs. The middle row shows a large inner radius (left) and a small inner radius (right) in the tori of recently dimmed and brightened AGNs, respectively. The bottom row shows the geometry of a slim accretion disk, where the torus clouds receive less radiation due to self-shadowing. The figure is not to scale.}
\label{fig:Schematic}
\end{figure*}

\textcolor{black}{\subsubsection{Possible Underlying Physical Mechanisms of the Observed Trend}}
Unlike the predicted slope of 0.5 by the dust sublimation model, our findings show a significantly shallower slope in the torus size-luminosity relation. This discrepancy can be primarily explained by anisotropic illumination from the accretion disk \citep[][]{Kawaguchi2010,Kawaguchi2011}, where radiation is stronger toward the poles and weaker in the equatorial plane due to the geometrically thin structure of the disk and the angular dependence of its emission. Limb darkening, which causes surface brightness to decline at higher viewing angles, also contributes to this anisotropy \citep[e.g.,][]{Netzer1987}. These effects arise from a combination of optical depth, vertical temperature gradients, and projection effects, making the observed emission angle-dependent. As a result, the dust sublimation radius is smaller in the equatorial plane and larger toward the poles. Since reverberation mapping studies primarily target type 1 AGNs viewed close to face-on, the observed lags tend to be shorter than the isotropic expectation. A common finding in many studies is that more luminous AGNs tend to have a geometrically thinner torus, meaning that the fraction of the geometrical thickness compared to the radius is smaller \citep[e.g.,][]{Ueda2003,Maiolino2007,Merloni2014}.
Thinner tori are associated with more rapid and potentially shorter (NIR) emission delays \citep[][]{Kawaguchi2011}. This effect, combined with the anisotropic illumination from the disk might contribute to the observed shallower slopes by causing a less pronounced increase in the lag with luminosity than predicted by a simple isotropic sublimation model and a geometrically constant torus. See the top row of Figure \ref{fig:Schematic} for a schematic illustration of these scenarios.

A second plausible explanation is that the inner dust torus in AGNs does not instantaneously adjust its size in response to changes in the luminosity of the central accretion disk \citep[e.g.,][]{Koshida2009,Pott2010,Oknyansky2014,Kokubo2020}. \textcolor{black}{The timescales for dust destruction and re-formation are likely different, with dust sublimation occurring relatively quickly following an increase in luminosity, while dust re-formation likely proceeds more slowly, resulting in an asymmetric response of the dust distribution radius \citep[e.g.,][]{Kishimoto2013}. If dusty clouds within the sublimation radius persist beyond typical variability timescales, the inner radius may remain temporarily small \citep[][]{Koshida2009}, possibly causing underestimation of its size in recently brightened AGNs (Figure \ref{fig:Schematic}, middle right panel). Conversely, after a decrease in luminosity, dust re-formation and cloud replenishment occur over longer timescales. Hence, in less luminous AGNs characterized by shorter variability timescales, the disk may dim while the inner radius of the torus remains larger than expected for the current luminosity (Figure \ref{fig:Schematic}, middle left panel). Thus, overestimated torus sizes in recently faded AGNs might be more frequently observed than underestimated ones in recently brightened AGNs, leading to a shallower slope than expected.}

The third possible explanation involves luminosity-dependent changes in the SED of AGNs, which may contribute to the observed deviation in the dust size-luminosity relation. The commonly assumed torus size luminosity scaling relation ($R_{\mathrm{tor}} \propto L^{0.5}$) presumes a fixed spectral shape for the illuminating radiation. However, the energy absorbed by dust grains depends on the spectral shape of the illumination, as the energy distribution across wavelengths in the SED varies \citep[e.g.,][]{Barvainis1987, Yoshii2014}. Observations by \citet{Krawczyk2013} reveal a luminosity-driven evolution in the UV/optical continuum, with more luminous AGNs exhibiting a bluer optical and redder far UV spectrum. {\citet{Xie2016} further supports this by showing that fainter AGNs tend to have redder UV spectra, possibly due to increased dust extinction. This luminosity-dependent SED implies that the UV-to-optical flux ratio has to be systematically lower for higher-luminosity AGNs. Since the dust sublimation radius is primarily determined by the UV flux, the dominant contributor to dust heating, using an optical luminosity as a proxy without correcting for the evolving SED shape can introduce a bias in estimating the torus radius. However, a similar argument has been proposed to explain the slope of $\sim$ 0.4 in the H$\beta$ BLR size-luminosity relation \citep[][]{Woo2024}. In that case, a tight correlation between optical and H$\beta$ luminosities suggests that optical luminosity remains a reasonable tracer of the ionizing flux. This casts some doubt on whether changes in the UV-to-optical ratio alone can fully account for the deviation from the expected $R_{\rm tor} \propto L^{0.5}$ scaling relation.

Recently, \cite{Chen2023} used the self-shadowing effect \citep[][]{Wang2014} to explain the shallow dust size-luminosity relation. Unlike thin disks, slim disks at high accretion rates are geometrically and optically thick in their inner regions \citep{Abramowicz1988}. A schematic view of slim disk is shown in bottom panel of Figure \ref{fig:Schematic}. Much energy is advected inward rather than radiated, reducing the direct UV/optical emission reaching the dust torus. As the dust sublimation radius depends on this illumination, it increases more slowly with luminosity than thin disk models predict. While this self-shadowing effect can occur across a range of AGN luminosities, it is more dominant in high-luminosity AGNs, which are more frequently associated with high Eddington ratios and, thus, more likely to host slim disks. As a result, the torus size-luminosity relation appears shallower than expected from standard disk models.
\textcolor{black}{While several physical scenarios may explain the shallower slope of the dust size-luminosity relation, we cannot currently quantify which, if any, of these models best describes our sample. A more definitive distinction would require detailed theoretical modeling and comparisons beyond the scope of this study.}

Finally, we note that the sublimation radius and its theoretical scaling are tied to very hot dust ($\sim$1500-1800 K) emitting in NIR. However, our W1 and W2-based torus size measurements trace cooler dust, suggesting that W1 and W2 band flux may not directly reflect the sublimation radius, potentially leading to deviations from theoretical expectations.

\subsection{Dependence of Torus Size on MIR Luminosity}

\begin{figure*}[htbp]
\centering
\includegraphics[width=0.92\textwidth]{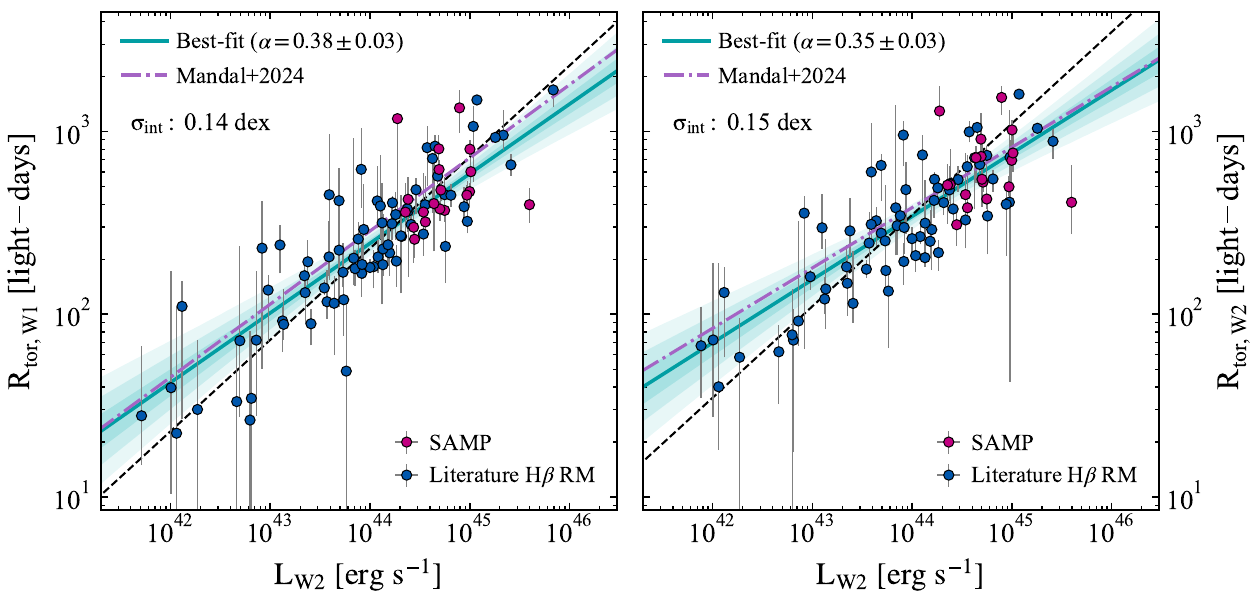}
\caption{Size of torus ($R_{\rm tor}$) from RM in the W1 band (left) and the W2 band (right) versus MIR W2 luminosity. The green solid line represents our best-fit regression with a free slope. The shaded region corresponds to the $1\sigma-3\sigma$ confidence intervals around the best-fit relation. The black dashed line shows the best fit with a fixed slope of 0.5. The purple dotted-dashed lines indicate regression results from \cite{Mandal2024}.}
\label{fig:size_lum_ir}
\end{figure*}

Given the shallower slope ($<0.5$) observed in the size-optical luminosity relation than the expected in Section \ref{sec:size-opt}, we investigate this relationship using W2 luminosity as an alternative, less obscured tracer of AGN-heated dust. The W2 band is more sensitive to the thermal emission from the inner torus and less contaminated by starlight and the AGN accretion disk than W1. Additionally, we correct our MIR data for accretion-disk contamination to more accurately isolate the thermal emission from the torus (see Section \ref{sec:wise_lc}). We fit the data using Equation \ref{equ:fit}, and find that the torus size scales with W2 luminosity as  $R_{\rm tor, W1} \propto L_{\rm W2} ^{0.38 \pm 0.03}$ and $ R_{\rm tor, W2} \propto L_{\rm W2}^{0.35 \pm 0.02}$, suggesting a tight correlation between the Torus size and the AGN MIR luminosity (see Table \ref{tab:regression_parameters} and Figure \ref{fig:size_lum_ir}). While our results are in good agreement with those of \cite{Mandal2024} (with $ R_{\rm tor, W1} \propto L_{\rm W1}^{0.40}$ and $ R_{\rm tor, W1} \propto L_{\rm W2}^{0.33}$), they deviate from the steeper relation reported by \cite{Minezaki2019} and \cite{Gravity2024} between dust size and MIR $12 \mu m$ luminosity as $R_{\rm tor} \propto L_{12 \mu m}^{\sim 0.5}$. This discrepancy could be attributed to the limited sample of $\sim$10 targets in these studies, which may impact the statistical robustness of their results.

\begin{table*}
\centering
\movetableright= -75mm
\caption{The results of time lag measurements for W1 and W2.}
\label{tab:final_lags}
\resizebox{\textwidth}{!}{
\fontsize{22pt}{22pt}\selectfont
\begin{tabular}{cccccccccccccccc}
\toprule
& & 
\multicolumn{8}{c}{\textbf{ICCF}} 
& \multicolumn{3}{c}{\textbf{MICA}} 
& \multicolumn{3}{c}{\textbf{H$\beta$}} \\
\cmidrule(lr){3-10} \cmidrule(lr){11-13} \cmidrule(lr){14-16}

Target Name & $z$
& $\tau_{\text{peak, W1}}$ & $\Delta\tau_{\text{peak, W1},-}$ & $\Delta\tau_{\text{peak, W1},+}$ 
& $\tau_{\text{cent, W1}}$ & $\Delta\tau_{\text{cent, W1},-}$ & $\Delta\tau_{\text{cent, W1},+}$ 
& $r_{\text{max, W1}}$ & $p(r_{\rm max})$ 
& $\tau_{\text{mica, W1}}$ & $\Delta\tau_{\text{mica, W1},-}$ & $\Delta\tau_{\text{mica, W1},+}$ 
& $\tau_{\text{H}\beta}$ & $\Delta\tau_{\text{H}\beta,-}$ & $\Delta\tau_{\text{H}\beta,+}$ \\

& 
& [day] &  & 
& [day] &  & 
& & 
& [day] & & 
& [day] & & \\
\\
(1)& (2) 
& (3) & (4) & (5)
& (6) & (7) & (8)
& (9) & (10)  
& (19) & (20) & (21) 
& (25) & (26) & (27)\\

\midrule
3C 120   & 0.033 & 215 & 43  & 127 & 228 & 37  & 125 & 0.76 & 0.010& 186 & 4  & 4  & 25.8 & 2.3 & 2.0 \\
3C 382   & 0.058 & 300 & 15  & 5   & 310 & 20  & 46  & 0.91 & 0.000 & 266 & 8  & 9  & 9.7  & 6.7 & 6.8 \\
3C390.3  & 0.056 & 285 & 117 & 33  & 267 & 50  & 31  & 0.91 & 0.000 & 199 & 9  & 8  & 18.5 & 2.3 & 3.4 \\
...  & ...   & ... & ... & ... & ... & ... & ... & ...  & ... & ... & ...& ...&  ...  & ... & ... \\
\bottomrule
\end{tabular}}
\raggedright
Note:
column (1) target name,
column (2) redshift,
columns (3)-(5) observed-frame peak W1-band time lag from ICCF, with its lower and upper uncertainties,
columns (6)-(8) observed-frame centroid W1-band time lag from ICCF, with its lower and upper uncertainties,
column (9) peak correlation coefficient from ICCF between optical and W1-band light curve,
columns (10) $p(r_{\rm max})$ from ICCF for W1,
columns (11)-(13) observed-frame W1-band time lag from MICA for W1, along with its lower and upper uncertainties,
columns (14)-(15) H$\beta$ lag measurement, with its lower and upper uncertainties. For clarity, the columns are renumbered in the printed version. Columns related to the W2 band [(11)–(18) and (22)–(24)] are omitted in the printed version. Only a portion of this table is shown here to demonstrate its form and content. A machine-readable version of the full table is available.
\end{table*}

\begin{table*}
    \centering
    \movetableright= -19mm
    \caption{The best-fitting regression parameters of the torus size-luminosity scaling relations.}
    \label{tab:regression_parameters}
    \begin{tabular}{llccccc}
        \toprule
        \multicolumn{2}{c}{Regression} & Slope ($\beta$) & Intercept ($\alpha$) & Intrinsic Scatter ($\sigma$) & Intercept$_{\rm fix}$& Intrinsic Scatter$_{\rm fix}$  \\
        \midrule
        \midrule
        \multicolumn{7}{l}{\textbf{W1 band}} \\
        \multirow{4}{*}{$R_{\rm tor}$ vs.} 
            & $L_{5100}$ & $0.35 \pm 0.03$ & $2.45 \pm 0.02$ & $0.15$ & $2.43 \pm 0.01$ & $0.18$ \\
            & $L_{5100}^{\rm merged}$ & $0.36 \pm 0.01$ & $2.44 \pm 0.01$ & $0.14$ & $2.35 \pm 0.01$ & $0.17$ \\
            & $L_{\rm bol}$ & $0.39 \pm 0.04$ & $2.09 \pm 0.04$ & $0.15$ & $1.98 \pm 0.01$ & $0.17$ \\
            & $L_{\rm W2}$ & $0.38 \pm 0.03$ & $2.39 \pm 0.02$ & $0.14$ & $2.36 \pm 0.01$ & $0.17$ \\
        \addlinespace
        \multicolumn{7}{l}{\textbf{W2 band}} \\
        \multirow{4}{*}{$R_{\rm tor}$ vs.} 
            & $L_{5100}$ & $0.33 \pm 0.03$ & $2.59 \pm 0.02$ & $0.16$ & $2.63 \pm 0.01$ & $0.21$ \\
            & $L_{5100}^{\rm merged}$ & $0.31 \pm 0.01$ & $2.57 \pm 0.01$ & $0.15$ & $2.49 \pm 0.01$ & $0.20$ \\
            & $L_{\rm bol}$ & $0.35 \pm 0.04$ & $2.26 \pm 0.05$ & $0.17$ & $2.18 \pm 0.01$ & $0.20$ \\
            & $L_{\rm W2}$ & $0.35 \pm 0.03$ & $2.53 \pm 0.02$ & $0.16$ & $2.54 \pm 0.01$ & $0.20$ \\
        \bottomrule
    \end{tabular}
    \noindent\parbox{\textwidth}{\footnotesize
\textbf{Note.} Rows with superscript $^{\mathrm{merged}}$ indicate regression results obtained from the merged sample, which combines our dataset with that of \cite{Mandal2024} after removing overlapping targets.  
The columns Intercept$_{\rm fix}$ and Intrinsic Scatter$_{\rm fix}$ indicate the best-fit parameters when the slope is fixed to 0.5.}
\end{table*}

\section{Discussion}
\label{sec:discussion}
\subsection{Deviation from the Dust Size--Luminosity Relation}
Although photoionization models predict a BLR size-luminosity relation of $R_{\rm BLR} \propto L_{\rm AGN}^{0.5}$, RM studies have shown that AGNs with high Eddington ratios ($\lambda_{\rm Edd} = L_{\rm bol}/L_{\rm Edd}$) often deviate from this trend, exhibiting systematically smaller BLR radii than expected \citep[e.g.,][]{Du2015,Du2016}. Several scenarios may cause this behavior, including self-shadowing effects in slim disks at high Eddington ratios \citep[][]{Wang2014}, nonlinear scaling between optical and ionizing emission \citep[e.g.,][]{Czerny2019, Fonseca2020}, and luminosity-dependent variations in the BLR ionization state \citep[e.g.,][]{Czerny2019, Fonseca2020}. For a more detailed discussion, see \citet[][]{Woo2024}.

Motivated by this, we will investigate the dependence of deviations from the dust size-luminosity relation on Eddington ratios. We divid our sample into three Eddington ratio bins as low-$\lambda_{\rm Edd}$ with log($\lambda_{\rm Edd}) \leq -1.5$, intermediate-$\lambda_{\rm Edd}$ with $-1.5 <$ log($\lambda_{\rm Edd}) < -0.5$ and high-$\lambda_{\rm Edd}$ with log($\lambda_{\rm Edd}) \geq -0.5$ (see Figure \ref{fig:eddington}). To quantify the deviation from the torus size-luminosity relation, we calculate $\rm \Delta log\ R_{\rm tor} = log\ (R_{\rm tor,obs}/R_{\rm tor,pred})$ for each band, where $R_{\rm Torus,obs}$ is the observed torus size and $R_{\rm Torus, pred}$ is the predicted value from the best-fit relation with a fixed slope ($\alpha_{\rm fix} = 0.5$). As shown in the right panels of Figure \ref{fig:eddington}, there is a negative trend between the deviation and Eddington ratio, with Spearman’s rank correlation coefficients $\rho = -0.186$ and $\rho = -0.289$ for W1 and W2, respectively. The blue shaded regions in the left panels highlight the highest Eddington-ratio bin, where most targets tend to have smaller torus sizes than predicted. These findings suggest that the dusty torus time lags in both W1 and W2 decrease with increasing Eddington ratio, \textcolor{black}{consistent with the self-shadowing scenario.} However, these trends appear only marginally significant, with $p$-values of $7.6 \times 10^{-2}$ for W1 and $7.3 \times 10^{-3}$ for W2. Note that when using our best fit with a free slope, this trend becomes weaker.

\begin{figure*}[ht!]
\centering
\includegraphics[width=0.92\textwidth]{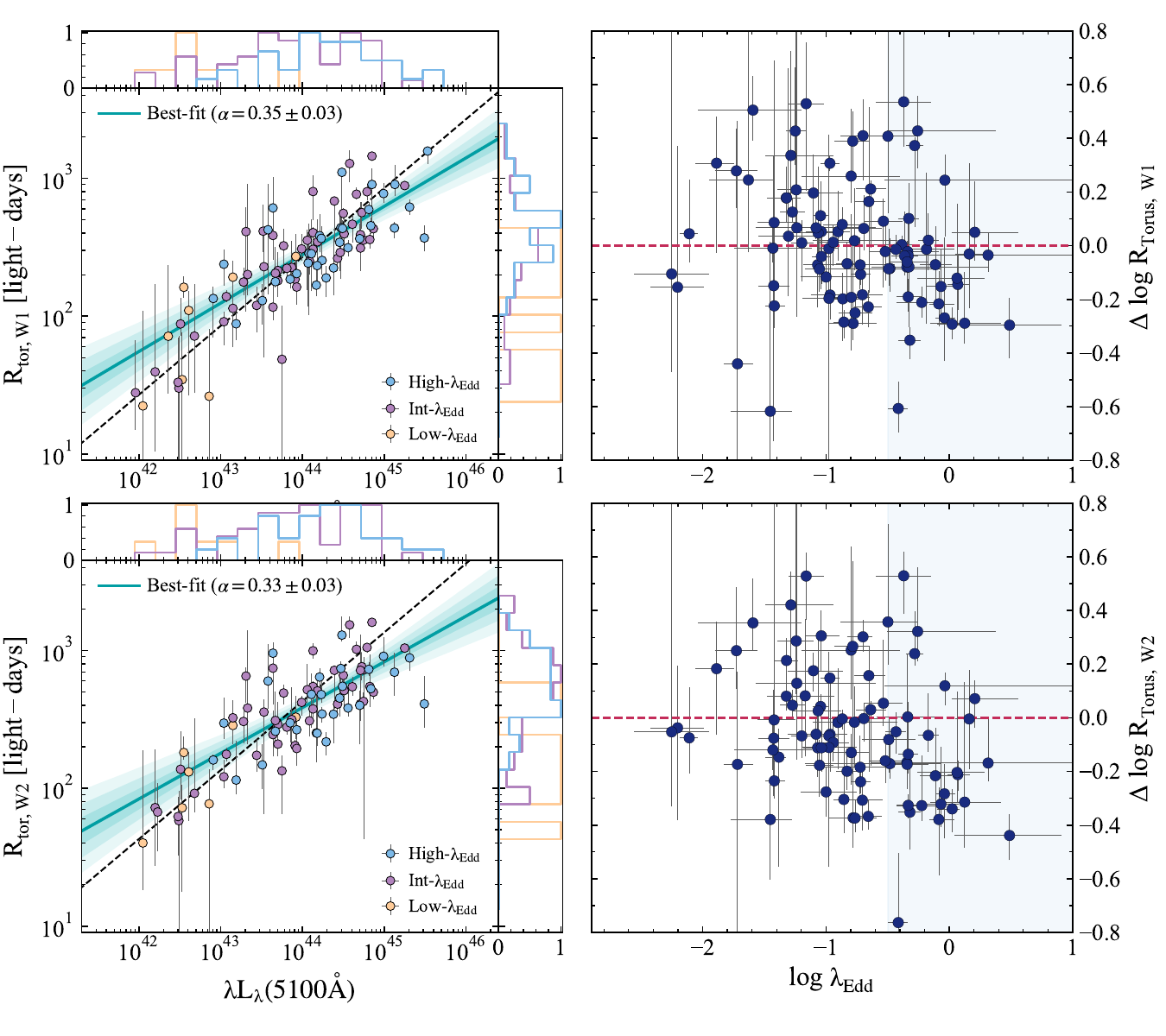}
\caption{The top row shows the relationship between $R_{\rm tor, W1}$ and $L_{5100}$ for three Eddington ratio bins in the W1 band (left) and the corresponding deviation values (defined as  $\rm \Delta log\ R_{\rm tor} = log\ (R_{\rm tor, obs}/R_{\rm tor, pred})$) as a function of the Eddington ratio (right). The shaded region indicates the highest bin of Eddington-ratio. The bottom row presents the same relationships for the W2 band. A negative trend is evident in the right panels, where deviation values decrease as the Eddington ratio increases for both bands.}
\label{fig:eddington}
\end{figure*}
  
Our findings align with \citet{Mandal2024} and \citet{Chen2023}, who reported {a reduced torus size} with increasing accretion rate {due to the slim disk self-shadowing effect.} In contrast, \citet{Yang2020} did not find a strong trend and proposed two selection effects to explain the slight negative correlation they observed.
First, they noted that the limited temporal coverage of RM data could be biased against detecting long lags due to reduced overlap between optical and MIR light curves. This would lead to systematically underestimated lag measurements, particularly for more luminous AGNs. {However, this effect may be less pronounced in our low-redshift ($z < 0.8$) sample.} For a typical luminous quasar at $z = 0.8$ with log $L_{5100} = 45.5\ erg\ s^{-1}$, the theoretical predicted dust lag is $\sim$6 years in the observed frame. Given the $\sim$20-year optical baseline and 14-year MIR baseline, $\sim$90\% of the full MIR light curve would still overlap with the earlier optical light curve, allowing us to reliably detect the lag.

Second, \citet{Yang2020} pointed out that at higher redshifts, the W1-band samples have shorter rest-frame IR wavelengths, which results in systematically shorter IR lags due to wavelength-dependent lag behavior. This effect could bias average lag measurements lower for higher-redshift and higher-luminosity AGNs. However, {in our recent work \cite[][]{Mandal2024}, we combined targets from \citet{Yang2020} with lower-redshift AGNs, remeasured the dust lags, and applied corrections for the wavelength dependence of the lag measurements, revealing a negative trend. In the present study, we apply the same wavelength correction described in Section \ref{sec:data_analysis} and find a consistent negative trend, supporting the result reported in our previous work. Therefore, we conclude that the observed negative correlation between dust size and the Eddington ratio is unlikely to be driven by selection effects.}

\subsection{Size Comparison of the BLR and Dusty Torus}

\begin{figure*}[ht!]
\centering
\includegraphics[width=0.92\textwidth]{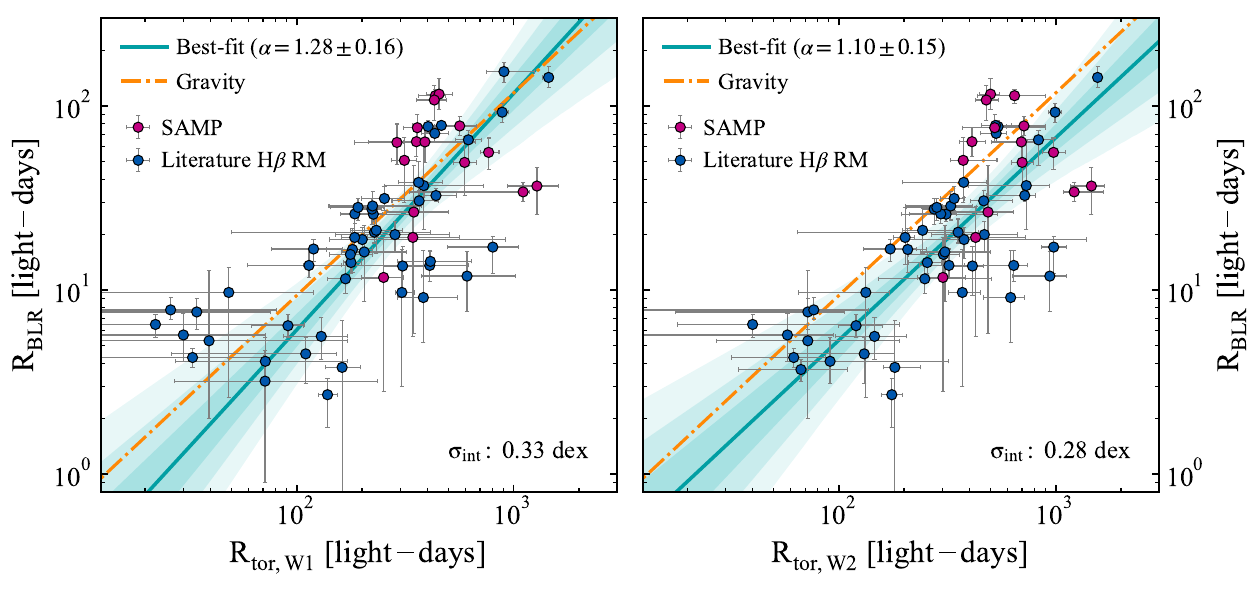}
\caption{The $\rm H\beta$ BLR size from \cite{Wang2024} plotted against the torus ($R_{\text{tor}}$) in the W1 band (left) and W2 band (right). The green solid line shows our best-fit regression and the shaded region corresponds to the $1-3\sigma$ confidence intervals around the best-fit line. The orange dot-dashed lines indicate results from optical/NIR interferometry presented by \cite{Gravity2023}.} 
\label{fig:Hbeta-size}
\end{figure*}

In this section, we compare the dust and BLR sizes for 67 AGNs with reliable dust lag and H$\beta$ lag measurements. We adopt the best H$\beta$ lags from \citet{Wang2024}. Figure \ref{fig:Hbeta-size} presents a comparison between the H$\beta$ BLR size and the torus sizes derived from W1 and W2 band lags. We obtain the best-fit relations as:

\begin{align}
\log \left( \frac{R_{\mathrm{BLR}}}{1\,\mathrm{light\text{-}day}} \right)
&= (1.397 \pm 0.043) + \notag \\
&\quad (1.28 \pm 0.16) \log \left( \frac{R_{\mathrm{tor,\;W1}}}{300\,\mathrm{light\text{-}days}} \right)
\end{align}
with an intrinsic scatter of $\sigma = 0.33$ dex for W1, and
\begin{align}
\log \left( \frac{R_{\mathrm{BLR}}}{1\,\mathrm{light\text{-}day}} \right)
&= (1.252 \pm 0.048) + \notag \\
&\quad (1.10 \pm 0.15) \log \left( \frac{R_{\mathrm{tor,\;W2}}}{300\,\mathrm{light\text{-}days}} \right)
\end{align}\\
with an intrinsic scatter of $\sigma = 0.28$ dex for W2. The pivot of 300 light days is close to the mean torus size in both W1 and W2 bands in our sample.}
These best-fit parameters offer an improved constraint compared to our previous work on 19 AGNs \cite[][]{Mandal2024}, which indicated a slope of $\sim$1.42 in both W1 and W2 bands. Our results also align well with the correlation $R_{\rm BLR} \propto R_{\rm tor}^{1.10}$ found by optical and NIR interferometric studies \citep[][illustrated by the orange dotted-dashed lines in Figure \ref{fig:Hbeta-size}]{Gravity2023} but show a slightly steeper trend than the near-unity slope reported by \cite{Chen2023} for 78 AGNs. Furthermore, our analysis shows that the torus size is $\sim$10 (14) times larger than the BLR for W1 (W2). These ratios are consistent with the estimates from \citet{Mandal2024}, who reported values of roughly 9.5 (W1) and 15.1 (W2) based on lag measurements. Earlier studies using K-band lags found smaller torus-to-BLR size ratios, typically ranging from about 4 to 7 \citep[e.g.,][]{Koshida2014,Gandhi2015,Du2015,Kokubo2020}. More recently, \citet{Chen2023} derived dust sizes using the MICA method, yielding ratios of 9.2 in W1 and 11.2 in W2, slightly lower than our values based on ICCF-derived BLR and dust lags. A larger sample with more accurately measured sizes is needed to better constrain this ratio and understand its potential dependence on AGN properties.

Both the BLR size-luminosity and torus size-luminosity relations exhibit similarly shallow slopes (both below $\sim$ 0.4), suggesting a common underlying physical mechanism. Given their close connection within the AGN structure, a near-linear correlation between BLR size and torus size is naturally expected. Importantly, torus sizes derived from MIR RM can serve as independent proxies for BLR size, offering estimates comparable to those based on optical luminosity (e.g., $L_{5100}$) and the BLR size-luminosity relation. This approach thus offers an alternative way to estimate black hole masses using similar scaling relations alongside traditional optical-based methods.

The upcoming Spectro-Photometer for the History of the Universe, Epoch of Reionization and Ices Explorer \citep[SPHEREx,][]{Dore2014} mission, operating for 2 years in the 0.75 to 5.0 $\mu$m range with frequent ecliptic pole observations, holds significant promise for AGN studies. Although its limited duration might preclude very long dust lag detections, the high cadence of observations will enable more precise time delay measurements for a larger sample, particularly crucial for probing the torus size-luminosity relation at lower luminosities where current data are sparse.

\section{Summary}
\label{sec:summary}
In this study, we employed a sample of 182 AGNs at $z < 0.8$ with uniform and reliable $\rm H\beta$ lag measurements from \cite{Wang2024} to investigate dust RM and the scaling relation of torus size. We constructed optical light curves using photometric data from ground-based surveys and compiled a complete WISE dataset up to its conclusion in August 2024. Our main findings are as follows:

\begin{enumerate}
\item Using both the ICCF and MICA methods, we found that the dust lag measurements from the two approaches are consistent, showing no significant systematic offset. Compared to previous studies, our results exhibit better agreement between the two methods, likely due to the application of more stringent selection criteria and the use of better-sampled light curves, made possible by a longer observational baseline.

\item We found that the torus size inferred from both W1 and W2 bands correlates strongly with the continuum luminosity at 5100 Å. However, the observed relations ($R_{\text{Torus, W1}} \propto L_{5100}^{0.35 \pm 0.03}$ for W1 and $R_{\text{Torus, W2}} \propto L_{5100}^{0.33 \pm 0.03}$ for W2) are shallower than the theoretical prediction of $R_{\text{tor}} \propto L^{0.5}$. After applying a nonlinear bolometric correction to convert $\rm L_{5100}$ to bolometric luminosity, we examined the relationship between torus size and bolometric luminosity, and found that the slopes remained nearly the same. This deviation may reflect effects such as anisotropic emission from the accretion disk combined with torus geometry or self-shadowing, among other possible factors.

\item As an alternative, we examined the correlation between torus size and AGN luminosity at W2, which is considered a less obscured tracer. We found that the resulting relations  ($R_{\rm tor, W1} \propto L_{\rm W2}^{0.38 \pm 0.03}$ for W1 and $R_{\rm tor, W2} \propto L_{\rm W2}^{0.35 \pm 0.03}$ for W2) remain shallower than the theoretical slope of 0.5 and closely resemble the trends seen with the 5100 Å continuum and bolometric luminosity.

\item We compared the torus size with that of the BLR and found that, on average, the torus is $\sim$10 (14) times larger than the BLR for W1 (W2). Our results indicate that the torus size-BLR size relation is slightly steeper than linear.

\end{enumerate}

\begin{acknowledgments}
This work has been supported by the Basic Science Research Program through the National Research Foundation of the Korean Government (grant No. 2021R1A2C3008486). A.K.M. acknowledges the support from the European Research Council (ERC) under the European Union’s Horizon 2020 research and innovation program (grant No. 951549).

\end{acknowledgments}

\bibliography{main}{}
\bibliographystyle{aasjournal}



\end{document}